\documentclass[pra,superscriptaddress,showpacs,twocolumn,aps]{revtex4-1}%
\usepackage{amsfonts,amssymb,amsmath,color}
\usepackage{graphics,graphicx,epsfig}
\usepackage{epstopdf}
\usepackage{amsmath}
\usepackage{amsfonts}
\usepackage{amssymb}
\usepackage{graphicx}%
\providecommand{\U}[1]{\protect\rule{.1in}{.1in}}
\newcommand{\be}{\begin{eqnarray}}
\newcommand{\ee}{\end{eqnarray}}
\begin{document}

\title{1-out-of-2 Oblivious transfer using flawed Bit-string quantum protocol}

\author{Martin Plesch}
\affiliation{Institute of Physics, Slovak Academy of Sciences, Bratislava, Slovakia}
\affiliation{Faculty of Informatics, Masaryk University,  Botanick\'a 68a, 602 00 Brno, Czech Republic}
\author{Marcin Paw\l{}owski}
\affiliation{Instytut Fizyki Teoretycznej i Astrofizyki, Uniwersytet
Gda\'nski, PL-80-952 Gda\'nsk, Poland}
\author{Matej Pivoluska}
\affiliation{Institute of Physics, Slovak Academy of Sciences, Bratislava, Slovakia}
\affiliation{Faculty of Informatics, Masaryk University,  Botanick\'a 68a, 602 00 Brno, Czech Republic}

\begin{abstract}
Oblivious transfer (OT) is an important tool in cryptography. It serves as a
subroutine to other complex procedures of both theoretical and practical
significance. Common attribute of OT protocols is that one party (Alice) has
to send a message to another party (Bob) and has to stay oblivious on whether
Bob did receive the message. Specific (OT) protocols vary by exact definition
of the task -- in the \emph{all-or-nothing} protocol Alice sends a single
bit-string message, which Bob is able to read only with $50\%$ probability,
whereas in \emph{1-out-of-2} OT protocol Bob reads one out of two messages
sent by Alice. These two flavours of protocol are known to be equivalent.
Recently a computationally secure all-or-nothing OT protocol based on quantum
states was developed in [A. Souto \textit{et. al.}, PRA \textbf{91}, 042306],
which however cannot be reduced to 1-out-of-2 OT protocol by standard means.
Here we present an elaborated reduction of this protocol which retains the
security of the original.

\end{abstract}
\maketitle

\section{Introduction}

Oblivious transfer (OT) is a very important building block for various
cryptographic protocols. Oblivious transfer exists in two flavours. The first
one, called \emph{all-or-nothing} OT provides a way to transfer a message from
Alice to Bob in such a way that Alice, as the sender, does not know whether
Bob did receive the message. More precisely, Alice sends an encoded message
$m$ to Bob. Bob can decode the message with probability $\frac{1}{2}$.
Importantly, Alice shall not learn whether Bob was able to read the message,
thus the name \emph{oblivious}.

It is a well-known fact that oblivious transfer cannot be executed with
unconditional security neither within the classical domain, nor in quantum
cryptography. Classically, the security is usually assured by computational
complexity arguments based on hardness of factoring (see \textit{e.g.}
\cite{Rabin}). This, however makes the existing protocols vulnerable against
attacks using quantum computers utilizing Shor's factoring algorithm
\cite{Shor}. This naturally motivated the research towards quantum algorithms
safe against quantum attacks, such as the result of Souto \textit{et. al.}
\cite{orgQOT}. There the authors presented a quantum computationaly secure
protocol for oblivious transfer under the assumption of at most few-qubits
measurements available.

Shortly afterwards He in \cite{comment} pointed out the incomplete security of
this protocol, mainly due to the fact that Alice can assure in certain runs
that Bob does not receive the message. Souto \textit{et. al.} replied
\cite{reply} by arguing that such a partially limited security of the protocol
is not of a significant hinder for its use. Although this statement might be
true for a subclass of utilizations of this oblivious transfer protocol, for
other classes this is certainly not the case. One of the later examples is its
possible utilization in a reduction to the \emph{1-out-of-2} oblivious
transfer protocol.

In 1-out-of-2 oblivious transfer, the task is slightly modified: Alice sends
two different messages $m_{0}$ and $m_{1}$. The aim of the protocol is to
assure that Bob is be able to read exactly one them, whereas Alice shall not
learn which message was accessed by Bob. As shown by the seminal work by
Cr\'epeau \cite{Crepeau}, all-or-nothing OT and 1-out-of-2 OT protocols are
equivalent in the sense that one can be efficiently used to implement the
other. This however only works if the starting protocol is perfectly secure,
which is not the case for the protocol introduced in \cite{orgQOT}. As
correctly pointed out in \cite{comment}, using the protocol of Souto \emph{et.
al.} together with Cr\'epeaus reduction leads to a complete loss of security.
This might have two possible causes -- either the protocol of \cite{orgQOT} is
unsuitable for reduction to 1-out-of-2 OT protocol per se, or the Cr\'epeaus
reduction is not appropriate for this flawed protocol. In this paper we show
the latter is the case -- we introduce an improved reduction that maintains
the level of security of the original quantum protocol throughout the
reduction. This reduction has however a more general use: it shows that the
two flavors of OT protocols are equivalent even in the presence of reasonably
bounded security flaw.

The paper is organized as follows: in the second section we briefly introduce
the protocol from \cite{orgQOT} as well as where its security flaw comes from.
In the third section we introduce a subroutine protocol called \emph{element
choosing protocol}, which allows a choice of an element from a set shared
between Alice and Bob in such a way that neither of them can influence the
choice significantly. Interestingly, this protocol is based on the flawed OT protocol
as well. The fourth section is devoted to the definition of the 1-out-of-2 OT
protocol itself and the fifth one to its security analysis. In the last
section we conclude our findings.

\section{Preliminaries}

\subsection{Quantum oblivious transfer of Souto \emph{et. al.}}

First we briefly analyze the protocol introduced in \cite{orgQOT}. Alice wants
to send a message, here a bit string $m$, to Bob, where Bob shall only with
probability $\frac{1}{2}$ learn the whole message, otherwise he shall not
learn (almost) anything. Analysis in \cite{orgQOT} correctly shows that

\begin{enumerate}
\item If both Alice and Bob are honest, Bob receives the message $m$ with
probability $\frac{1}{2}$

\item Bob knows whether he got the message or not

\item Bob's cheating is limited to a negligible probability

\item Alice does not know whether Bob received the message if she was sending
it honestly.
\end{enumerate}

However, as correctly pointed out by He in \cite{comment}, Alice can perform a
more sophisticated attack. She can decide to use incorrect message in the last
step of the protocol, which causes Bob's failure to obtain the correct message.

Bob cannot check whether Alice was honest in a single run of the protocol. On
the other hand, Alice cannot increase the probability of Bob to obtain the
correct message above $1/2$. Thus, every run of the protocol in which Alice
decides to cheat causes a decrease of the frequency of Bob's successes below
$50\%$. Thus, Alice and Bob can agree on a constant $s$ being a security
parameter. With $N$ repetitions of the protocol, Bob will terminate the
cooperation with Alice if he receives less than $\frac{N}{2}-\beta\sqrt{N}$
messages from Alice. If Alice wants to avoid termination of the protocol, she
can act dishonestly only in up to $\frac{o(\beta)}{\sqrt{N}}$ fraction of
rounds, which can be made arbitrary small with increasing $N$. As shown in
\cite{reply}, this is fine for some applications of the protocol, but
certainly not for all. Next we show why the reduction to 1-out-of-2 OT
protocol due to Cr\'{e}peau \cite{Crepeau} fails.

\subsection{Failure of Cr\'{e}peaus reduction}

The original reduction from all-or-nothing to 1-out-of-2 oblivious transfer
protocol from \cite{Crepeau} works as follows:

\begin{enumerate}
\item Alice chooses at random $N$ bits $r_{1}, r_{2}, ..., r_{N}$.

\item For each of these $N$ bits, Alice uses the all-or-nothing OT protocol to
disclose the bit $r_{i}$ to Bob.

\item Bob selects indices $U=\{i_{1},i_{2},...,i_{n}\}$ and $V=\{i_{n+1}%
,i_{n+2},...,i_{2n}\}$ where $n=\frac{N}{3}$ with $U\cap V=\emptyset$.
Additionally, it is required that he knows $r_{i_{l}}$ for each index $i_{l}\in U$.
If he didn't receive enough messages to select $U$, the protocol is terminated.

\item Bob sends $(X,Y)=(U,V)$ if he wants to read $m_{0}$ and $(X,Y)=(V,U)$ otherwise.

\item Alice computes $k_{0} = \bigoplus_{x\in X}r_{x}$ and $k_{1} =
\bigoplus_{y\in Y}r_{y}$.

\item Alice returns to Bob $k_{0}\oplus m_{0}$ and $k_{1}\oplus m_{1}$.

\item Bob computes $\bigoplus_{u\in U}r_{u}\in\{k_{0},k_{1}\}$ and uses it to
get his secret bit $m_{0}$ or $m_{1}$ according to his previous choice.
\end{enumerate}

The main idea behind this reduction is hidden in the step $3$. If number $N$
of the messages sent to Bob is large, then it is highly improbable that he
will receive less than $\frac{N}{3}$ of the messages correctly. On the other
hand, it is equally improbable that he will receive more than $\frac{2N}{3}$
of the messages correctly. Therefore, except for a marginal probability, he
will be able to produce a subset $U$ consisting of one third of rounds for
which he knows all the messages, but not both subsets $U$ and $V$. Later Alice
encrypts two bits: one using the messages from $U$ as the key, the second
using messages from $V$, knowing that Bob will be able to decrypt only one. If
the underlying all-or-nothing OT is not compromised, Alice will not know which
bit Bob can decrypt, as she has no information on which of the two sets
consist of messages Bob has successfully received.

However, as it was pointed out in \cite{comment}, the situation changes
drastically if Alice can cheat, even to a small extent, as in protocol of
\cite{orgQOT}. Alice can choose $s$ rounds where she knows Bob did not receive
the correct message. When Bob chooses rounds for the subset $V$, he can choose
roughly $\frac{N}{6}$ indices of received messages (Bob received roughly
$\frac{N}{2}$ messages, however he needs $\frac{N}{3}$ of them for $U$), but
needs to select another roughly $\frac{N}{6}$ indices belonging to the
messages he did not receive. Note that each out of the $s$ dishonest rounds
will be chosen with probability at least $\frac{1}{3}$. The probability that
Bob chooses at least one of these rounds as a member of $V$ is thus can be
upper bounded as $1-\left(  \frac{2}{3}\right)  ^{s}$, which quickly
approaches unity with increasing $s$.

If there is at least a single element from $s$ in $V$, Alice can with
certainty learn whether $(X,Y)=(U,V)$ or $(X,Y)=(V,U)$. Thus she can learn
which bit Bob intents to read with only negligible probability to be caught.

\subsection{Security Parameters}

In what follows we analyse the security of different two different protocols
implemented with the help of flawed OT protocol of \cite{orgQOT}, where the
first analysed protocol is used as a subroutine for the second one. We are in
principle interested in three basic parameters: the probability of the protocols
to fail if both parties are honest, denoted $p_{f}$, and the probabilities that
Alice or Bob cheat successfully, $p_{A}$ and $p_{B}$ respectively (with the
other party being honest). In what follows we evaluate the parameters of the
protocol depending on these three output parameters. To make the analysis
easier to access, we set $p_{A}=p_{B}=\varepsilon$ and $p_{f}=\frac{1}{2}$, to
get a single security parameter of the respective protocol.

\section{Element choosing protocol}

First we introduce a subroutine protocol which uses a possibly compromised
all-or-nothing OT as a primitive and allows the parties to choose one element
from some large set $\mathcal{T}$ with cardinality $N_\mathcal{T}$. This set has two
subsets $\mathcal{A}\subset\mathcal{T}$ and $\mathcal{B}\subset\mathcal{T},
\mathcal{A}\cap\mathcal{B} = \emptyset$
with $N_\mathcal{A}$ and $N_\mathcal{B}$ elements respectively. Alice wins if an element from
$\mathcal{A}$ is chosen and Bob wins if an element from $\mathcal{B}$ is
chosen. Our goal is to have a protocol in which, with high probability,
neither of them wins, thus the chosen element does not belong to either
$\mathcal{A}$ or $\mathcal{B}$.

The protocol is defined as follows:

\begin{enumerate}
\item Alice and Bob agree on a parameter $\alpha<\frac{1}{2}$.

\item Parties count the elements of $\mathcal{T}$ and label them with integers
from $1$ to $N_{\mathcal{T}}$.

\item Alice chooses at random $N_\mathcal{T}$ messages $r_{1},r_{2},...,r_{N_\mathcal{T}}$,
where each of the messages consists of $\ell$ bits. Alice will choose $\ell$
large enough so that the probability of guessing $r_{i}$ will be low enough in
comparison to any other probability within the protocol.

\item Alice sends each message $r_{i}$ with the all-or-nothing OT protocol
from \cite{orgQOT}.

\item Bob replies by publishing $\alpha N_\mathcal{T}$ messages he received correctly.
If he is not able to do so, the protocol is aborted.

\item Indices of the published $\alpha N_\mathcal{T}$ messages are used as the new set $\mathcal{T}$.
Alice and Bob repeat points 2-5 of this protocol $x$ times, in each round
keeping an $\alpha$ fraction of previously held messages; they will end up
with $N=\alpha^{x}N_\mathcal{T}$ messages.

\item Alice randomly chooses one of these messages to be the outcome of the protocol.
\end{enumerate}

The protocol aborts if Bob announces incorrectly any received message (which shall not happen
for honest players) or if Bob is not able to announce $\alpha N_\mathcal{T}$ messages
in some round. At the end of the protocol, Alice and Bob choose a single
message $r_{i}$.

\subsection{The probability of failure}

If the players are honest, the probability of failure in the first round
$p_{f,1}$ is bounded from above by
\begin{equation}
p_{f,1}\leq e^{-\frac{(1-2\alpha)^{2}N_\mathcal{T}}{4}},
\end{equation}
where we used the Chernoff bound for $\Pr(X\leq\alpha N_\mathcal{T})$. To further
simplify the notation we introduce a new symbol $\xi=\frac{1}{2}-\alpha>0$,
leading to a failure probability of
\begin{equation}
p_{f,1}\leq e^{-\xi^{2}N_\mathcal{T}}.
\end{equation}
Recall that after the first stage, Alice and Bob will keep $\alpha N_\mathcal{T}$
indices from the set $\mathcal{T}$.

After $x$ rounds, the probability of failure of the protocol will be upper
bounded by the sum of probability of failures in each respective run
\begin{align}
p_{f}  &  \leq{\displaystyle\sum\limits_{i=1}^{x}} p_{f,i}= {\displaystyle\sum
\limits_{i=1}^{x}} e^{-\xi^{2}N_\mathcal{T}\alpha^{i-1}}=\label{pf_ech}\\
&  = {\displaystyle\sum\limits_{j=1}^{x}} e^{-\xi^{2}N\alpha^{-j}}<
{\displaystyle\sum\limits_{j=1}^{x}} e^{-\xi^{2}N2^{j}}<\nonumber\\
&  <{\displaystyle\sum\limits_{j=1}^{x}} e^{-2j\xi^{2}N}< {\displaystyle\sum
\limits_{j=1}^{\infty}} e^{-2j\xi^{2}N}<\nonumber\\
& <e^{-2\xi^{2}N}\left(  1-e^{-2\xi^{2}N}\right)  ^{-1} <2e^{-2\xi^{2}%
N},\nonumber
\end{align}
where we used $e^{-2\xi^{2}N}<\frac{1}{2}$, which is a reasonable expectation,
otherwise the protocol would fail with high probability already in the first round.

\subsection{Bob's cheating}

The probability of selecting an index from $\mathcal{B}$  in the last round (after $x$ round) is given by $\frac
{N_\mathcal{B}^{x}}{\alpha^{x}N_\mathcal{T}}$, where $N_\mathcal{B}^{x}$ is the number of messages
left from $\mathcal{B}$ in the last step of the protocol. We can neglect the
probability that Bob can pretend to get a message he did not receive (which
can be achieved by choosing a sufficiently high $\ell$). Then the number of
Bob's winning messages is halved on average in each round, as the probability
of each message to reach Bob is $\frac{1}{2}$. Thus, the cheating probability
of Bob is bounded as
\begin{equation}
p_{B}\leq\frac{N_\mathcal{B}}{(2\alpha)^{x}N_\mathcal{T}}. \label{pb_ech}%
\end{equation}
Bob has in fact no possibility of active cheating.

\subsection{Alice's cheating}

Alice can cheat, if there is at least one message from $\mathcal{A}$ left at
the end of the protocol, by simply choosing this message. The probability of
this happening, if she is not actively cheating in the previous runs of the
protocol is $\frac{N_\mathcal{A}2^{-x}}{N}$. However, in each round she can reduce
the number of messages she sends to Bob honestly. In this way she can make
sure that some messages $r_{i},i\notin\mathcal{A}$ are not received by Bob,
thus increasing the chance that Bob will choose some of the received messages
from $\mathcal{A}$ for the next round. Every such attempt increases the
probability that the protocol fails. Here we upper bound the probability that
Alice can cheat by showing that either her cheating probability, or the
success probability of the whole protocol, are strictly upper bounded.

Let us divide the possible strategies of Alice in each round into two classes.
The first class contains strategies where she sends less than $2\alpha$
fraction of all the messages in the given round honestly -- we call these
strategies \textit{hard} and the remaining ones we call \textit{soft}. In a
hard strategy, in principle, Alice can send honestly only messages for indices in $\mathcal{A}$,
however, provided that $\mathcal{A}$ is small enough, the probability that Bob will
correctly receive enough messages would be extremely small.

For each round in which Alice uses a hard strategy, the failure probability of
the protocol will be multiplied by $\frac{1}{2}$ (or higher) due to the fact
that Bob will receive enough messages for a reply with probability less than
$\frac{1}{2}$. For each round of a soft strategy, at least half of the
messages will not be received by Bob, so the size of the Alice's winning set
will be reduced by at least $\frac{1}{2}$. Therefore with $i$ rounds of hard
strategy and $x-i$ rounds of a soft one, the protocol successfully passes with
$p_{pass}\leq2^{-i}$ and if it passes, Alice can cheat with probability
$p_{s,A}\leq$ $N_\mathcal{A}2^{-x+i}$. So in total the probability for Alice to
successfully cheat is bounded by
\begin{equation}
p_{A}\leq N_\mathcal{A}2^{-x}.\label{pa_ech}%
\end{equation}

It is clear that we cannot always choose $\alpha$ and $x$ in such a way that
both failure and cheating probabilities will be low. As this protocol is only
a subroutine, we will have to carefully choose parameters $N_\mathcal{T}$, $N_\mathcal{A}$ and
$N_\mathcal{B}$ in order to allow the protocol to work correctly.

\section{1-out-of-2 OT protocol}

Now we are ready to formulate the main protocol. In the protocol both $m_{0}$
and $m_{1}$ are single bit messages denoted $b_{0}$ and $b_{1}$. Let us denote $B$ the choice of the bit wished by Bob  --
if he wants to learn $b_0$, he chooses $B=0$, otherwise he chooses $B=1$.

\begin{enumerate}
\item Alice and Bob agree on the security parameters $c$ and $\beta$.

\item Alice first splits her messages $b_{0}$ and $b_{1}$ into groups of
random bits $b_{0}^{j}$ and $b_{1}^{j}$ such that $\bigoplus_{j=1}^{c}%
b_{0}^{j}=b_{0}$ and $\bigoplus_{j=1}^{c}b_{1}^{j}=b_{1}$.

\item For every pair $b_{0}^{j}$ and $b_{1}^{j}$, Alice chooses $N$ single bit
messages $r_{1},r_{2},...,r_{N}$.

\item For each of these $N$ messages, Alice uses the all-or-nothing OT
protocol to disclose them to Bob. The protocol is terminated if Bob did not
receive at least $n=\frac{N}{2}-\beta\frac{\sqrt{N}}{2}$ messages.

\item Let us denote $S$ the set of indices of messages that Bob received correctly. Bob
chooses $n$ pairs of indices $\{(u_{i},v_{i})\}_{i=1}^{n}$, such that $\forall i, u_i\in S$ if $B=0$, otherwise
he chooses $\forall i, v_i\in S$.
Also, for each pair Bob chooses a single random bit $k_i$. If $k_i = 1$, he switches the
order of the pair $(u_i,v_i)$, otherwise he keeps the order intact. After this operation,
Bob publishes the set of pairs, but keeps $k_i$ secret.

\item Alice and Bob use the element choosing protocol described above with
$\mathcal{T}$ being the set of pairs of indices announced by Bob. Therefore,
$N_\mathcal{T}=n=\frac{N}{2}-\beta\frac{\sqrt{N}}{2}$. Let us denote the pair chosen $h$.

\item Bob announces the bit $k_h$ so that Alice can switch the order of the selected pair if $k_h=1$;
let us denote the final key $(u,v)$.

\item Alice sends $b_{0}^{j}\oplus r_{u}$ and $b_{1}^{j}\oplus r_{v}$
to Bob. Since he knows one of the messages $r_{u}$ or $r_{v}$,
he can calculate $b_{B}^{j}$ of his choice.

\item Steps 2-6 are repeated $c$ times for all pairs of $b_{0}^{j}$ and
$b_{1}^{j}$. To obtain bit $b_{B}$ in which Bob is interested, he needs to
learn all the bits $b_{B}^{j}$.
\end{enumerate}

\subsection{Probability of failure}

The protocol can fail in two different parts. First, in the step 4, Bob might
receive less than $n$ messages correctly. For honest parties this probability
is bounded by $e^{-\beta^{2}/2}$ . In step 6 the element choosing protocol
might fail as well, with the probability (\ref{pf_ech}) derived in previous
section. Plugging in $N_\mathcal{T}=n$ we get $p_{f}<$ $2e^{-2\xi^{2}\alpha^{x}N_\mathcal{T}%
}=2e^{-2\xi^{2}\alpha^{x}\left(  \frac{N}{2}-\beta\frac{\sqrt{N}}{2}\right)
}$, recalling $\xi=\frac{1}{2}-\alpha$. So the probability of failure in any
particular round $j$ of the protocol is bounded by
\begin{align}
p_{f}^{j} &  <e^{-\beta^{2}/2}+2e^{-2\xi^{2}\alpha^{x}\left(  \frac{N}%
{2}-\beta\frac{\sqrt{N}}{2}\right)  }\\
&  <e^{-\beta^{2}/2}+2e^{-\xi^{2}\alpha^{x}\frac{N}{2}}\nonumber
\end{align}
for
\begin{equation}
\beta\leq\frac{\sqrt{N}}{2},\label{beta_bound1}%
\end{equation}
what is a condition on the parameters which will be taken into account by
choosing security parameters of the protocol. Further we can bound %

\begin{equation}
p_{f}^{j}<e^{-\beta^{2}/2}+2e^{-N\xi^{2}2^{-x-1}\left(  1-x\xi\right)  }%
\end{equation}
using $\alpha^{x}=\left(  \frac{1}{2}-\xi\right)  ^{x}<\frac{1}{2^{x}}\left(
1-x\xi\right)  $. Using another condition on the parameters%
\begin{equation}
\xi\leq\frac{1}{2x}\label{dzeta_bound}%
\end{equation}
we get%

\begin{equation}
p_{f}^{j}<e^{-\beta^{2}/2}+2e^{-N\xi^{2}2^{-x-2}}.\label{pf_j}%
\end{equation}
Thus the final probability to abort an honest protocol is bounded by the sum
of probability of failure for individual rounds reading
\begin{equation}
p_{f}<c\left(  e^{-\beta^{2}/2}+2e^{-N\xi^{2}2^{-x-2}}\right)  .\label{pf_st}%
\end{equation}

\subsection{Bob's cheating}

It is fairly hard for Bob to successfully cheat in the protocol. To obtain
both bits $b_{0}$ and $b_{1}$, he has to learn both bits $b_{0}^{j}$ and
$b_{1}^{j}$ in all $c$ rounds of the main protocol. Let us here analyze a
single round $j$.

From honest Alice, Bob will successfully receive on average $\frac{N}{2}$
messages, thus the probability that he receives more than $\frac{N}{2}%
+\beta\frac{\sqrt{N}}{2}$ is less than $e^{-\beta}$. So with high probability, he can produce no more than $N_\mathcal{B}=\frac{N}%
{4}+\beta\frac{\sqrt{N}}{4}$ pairs of indices,  where he received both
corresponding messages. Now he has to force the element choosing protocol to
select one of these pairs, which happens with a probability (\ref{pb_ech})
\begin{align}
p_{B}^{j}  &  =\frac{N_\mathcal{B}2^{-x}}{N_\mathcal{T}\alpha^{x}}\label{pb_j}\\
&  \leq\left(  2\alpha\right)  ^{-x}\frac{\frac{N}{4}+\beta\frac{\sqrt{N}}{4}%
}{\frac{N}{2}-\beta\frac{\sqrt{N}}{2}}\nonumber\\
&  <\left(  2\alpha\right)  ^{-x}\frac{1}{2}\frac{1+\frac{\beta}{\sqrt{N}}%
}{1-\frac{\beta}{\sqrt{N}}}\nonumber\\
&  \leq\left(  1-2\xi\right)  ^{-x}\frac{2}{3}\nonumber
\end{align}

for
\begin{equation}
\beta\leq\frac{\sqrt{N}}{5},\label{beta_bound2}%
\end{equation}
what is just a more strict limitation on $\beta$ comparing to
(\ref{beta_bound1}). Probability of cheating in all rounds of the protocol is
thus limited to
\begin{equation}
p_{B}<\left(  1-2\xi\right)  ^{-cx}\left(  \frac{2}{3}\right)  ^{c}%
.\label{pb_st}%
\end{equation}

\subsection{Alice's cheating}

For Alice the cheating is seemingly easier. It is enough for her to learn the
choice of Bob in a single round of the protocol. She can do it by sending a
portion of messages that are unreadable to Bob. If one of these messages
appears as a pair $(u,v)$ in one of the $c$ instances of the element choosing protocol
in step 6, Alice learns that the
corresponding bit is not the one that Bob is interested in.

Similarly to the case of Bob, Alice can send in each round of the protocol not
more than $\beta\sqrt{N}$ garbage messages, otherwise the success probability
will be lower than $e^{-\beta}$. She wins if she
selects in the Element choosing protocol a pair where one of her garbage
messages is present, what happens with probability $p_{A}^{j}\leq\beta\sqrt
{N}2^{-x}$. The total probability of cheating is upper bounded by the sum of
probabilities in each round%
\begin{equation}
p_{A}<c\beta\sqrt{N}2^{-x}. \label{pa_st}%
\end{equation}

\section{Security analysis}

Let us now analyze how to achieve security of the whole protocol. We will
define the protocol as secure, if the probabilities of cheating of both Bob
and Alice will be bounded by a given constant $\varepsilon$%
\begin{align}
p_{A}  &  <c\beta\sqrt{N}2^{-x}\leq\varepsilon,\label{pa}\\
p_{B}  &  \leq\left(  1-2\xi\right)  ^{-cx}\left(  \frac{2}{3}\right)
^{c}\leq\varepsilon. \label{pb}%
\end{align}
At the same time we expect the protocol to have a reasonable chance to finish
successfully if the parties are honest, hence we expect
\begin{equation}
p_{f}<c\left(  e^{-\beta^{2}/2}+2e^{-N\xi^{2}2^{-x-2}}\right)  \leq\frac{1}%
{2}. \label{pf}%
\end{equation}

In the following paragraphs we will analyze these conditions with the aim of finding
a suitable choice of parameters that fulfill them. Equation (\ref{pf}) is satisfied if both%
\begin{align}
ce^{-\beta^{2}/2} &  =\frac{1}{4}\label{pf1}\\
ce^{-N\xi^{2}2^{-x-2}} &  =\frac{1}{8}.\label{pf2}%
\end{align}
From (\ref{pf1}) we can set
\begin{equation}
\beta=\sqrt{2}\ln^{1/2}\left(  4c\right)  \label{beta1}%
\end{equation}
and from (\ref{pf2}) we get
\begin{equation}
N=\ln\left(  8c\right)  \xi^{-2}2^{x+2}.\label{K1}%
\end{equation}
Plugging these into (\ref{pa}) we get%
\begin{equation}
c\ln^{1/2}\left(  8c\right)  \ln^{1/2}\left(  4c\right)  \xi^{-1}%
2^{3/2-x/2}\leq\varepsilon,
\end{equation}
which is satisfied if
\begin{equation}
\xi=c\ln\left(  8c\right)  \varepsilon^{-1}2^{3/2-x/2},\label{dzeta1}%
\end{equation}
where we used $\ln\left(  8c\right)  >\ln\left(  4c\right)  $. This can be
further plugged into (\ref{pb})%
\begin{equation}
\left[  \left(  \frac{2}{3}\right)  \left(  1-c\ln\left(  8c\right)
\varepsilon^{-1}2^{5/2-x/2}\right)  ^{-x}\right]  ^{c}\leq\varepsilon
.\label{LogBound}%
\end{equation}
Let us now analyze (\ref{LogBound}) in more detail. A part of this inequality
has the form
\begin{equation}
\left(  1-c\ln\left(  8c\right)  \varepsilon^{-1}2^{5/2-x/2}\right)  ^{-x}.
\end{equation}
Let us now, for a fixed choice of $c$ and $\varepsilon$, implicitly define
$x_{0}$ by
\begin{equation}
c\ln\left(  8c\right)  \varepsilon^{-1}2^{5/2-x_{0}/2}=\frac{1}{4x_{0}%
}.\label{x0}%
\end{equation}
Clearly, for fixed $c$ and $\varepsilon$ all $x\geq x_{0}$%
\begin{equation}
c\ln\left(  8c\right)  \varepsilon^{-1}2^{5/2-x/2}\leq\frac{1}{4x}.
\end{equation}
This is satisfied for all $x\geq x_{0}$ such that%
\begin{equation}
\left[  \left(  \frac{2}{3}\right)  \left(  1-\frac{1}{4x}\right)
^{-x}\right]  ^{c}\leq\varepsilon.
\end{equation}
Further we know that for all $x\geq1$%
\begin{equation}
e^{1/4}\leq\left(  1-\frac{1}{4x}\right)  ^{-x}\leq\frac{4}{3}.
\end{equation}
Using this we can state that (\ref{LogBound}) will be satisfied for all $x>1$
if
\begin{equation}
\left(  \frac{8}{9}\right)  ^{c}\leq\varepsilon.
\end{equation}
Hence we can choose
\begin{equation}
c=\frac{\ln(\varepsilon)}{\ln\left(  8/9\right)  }.\label{c}%
\end{equation}
Plugging (\ref{c}) back to (\ref{x0}) we get an implicit definition of $x$
\begin{equation}
\frac{1}{\varepsilon}\frac{\ln(\varepsilon)}{\ln\left(  8/9\right)  }%
\ln\left(  8\frac{\ln(\varepsilon)}{\ln\left(  8/9\right)  }\right)
2^{5/2-x/2}=\frac{1}{4x},\label{x}%
\end{equation}
which can be efficiently solved numerically. We can further plug (\ref{c})
into (\ref{beta1}) and get
\begin{equation}
\beta=\sqrt{2}\ln^{1/2}\left(  4\frac{\ln(\varepsilon)}{\ln\left(  8/9\right)
}\right)  .\label{beta}%
\end{equation}
We also can use this result and (\ref{dzeta1}) to define $\xi$%
\begin{equation}
\xi=\frac{1}{\varepsilon}\frac{\ln(\varepsilon)}{\ln\left(  8/9\right)  }%
\ln\left(  8\frac{\ln(\varepsilon)}{\ln\left(  8/9\right)  }\right)
2^{3/2-x/2},\label{dzeta}%
\end{equation}
which involves the numerical solution for $x$. For obtaining a simpler formula
we utilize (\ref{x}) again and get
\begin{equation}
\xi=\frac{1}{8x},\label{dzeta_simplified}%
\end{equation}
thus fulfilling the condition (\ref{dzeta_bound}). And finally we get the
solution for $N$ by plugging everything into (\ref{K1})%
\begin{equation}
N=\ln\left(  8\frac{\ln(\varepsilon)}{\ln\left(  8/9\right)  }\right)
x^{2}2^{2x+8}.\label{K}%
\end{equation}
Comparing (\ref{beta}) and (\ref{K}) it is obvious that (\ref{beta_bound2}) is
satisfied for all $x>1$.

This concludes the proof of the security of the protocol -- for a given
$\varepsilon$, we can choose parameters of the protocol according to
(\ref{c}-\ref{K}) to be sure that the failure probability will not be larger
than $\frac{1}{2}$ and cheating of both Alice and Bob will be bounded by
$\varepsilon$. It is important to stress here that this choice is a mere proof
of existence of suitable set of parameters of the protocol -- if we would be
interested in efficient values, numerical search within the constrains
(\ref{pa}-\ref{pf}) would certainly be more appropriate.

\section{Conclusions}

Oblivious transfer protocols are important building blocks in cryptography.
Perfect all-or nothing OT protocol can be used to construct a 1-out-of-2 OT
protocol (and vice-versa), but this is in general not true for protocols with
security security flaw as the one in \cite{orgQOT}. And as unconditionally
secure OT protocols provably do not exist, it is of utmost importance to
investigate protocols with security based on reasonable assumptions.

In this paper we have introduced an improved version of the reduction protocol
from all-or-nothing to 1-out-of-2 OT protocol. Contrary to the existing
results, this reduction is immune against a security loophole in the original
protocol allowing Alice to learn partial one-sided information in a small
fraction of the protocol runs. This, in particular, allows the use of the
protocol suggested in \cite{orgQOT} for such a reduction. But our result is
more general, opening the possibility for utilizing other imperfect
implementations of all-or-nothing OT protocols as well.

\section*{Acknowledgments}

MPl and MPi were supported by the project VEGA 2/0043/15. MPl was supported by the Czech Science Foundation GA\v{C}R project
P202/12/1142 and MPi by the SAS Schwarz fund. MPa acknowledges the support
form NCN grant no. 2014/14/E/ST2/00020.

\end{document}